\documentclass[12pt]{article}

\usepackage{latexsym}
\usepackage{amssymb,amsfonts,amsmath}
\usepackage{bbm}
\usepackage{amssymb}
\usepackage{hyperref}
\usepackage{amsfonts}
\usepackage{dsfont}
\usepackage{slashed, tensor}
\usepackage{booktabs}

\parskip=\medskipamount

\newcommand{\be}{\begin{equation}}
\newcommand{\ee}{\end{equation}}
\newcommand{\bea}{\begin{eqnarray}}
\newcommand{\eea}{\end{eqnarray}}

\newcommand{\ba}{\begin{array}}
\newcommand{\ea}{\end{array}}
\newcommand{\cN}{{\cal N}}

\numberwithin{equation}{section}

\begin{document}

\begin{titlepage}
\begin{flushright}
January 2019 \\
\end{flushright}
\vspace{5mm}

\begin{center}
{\Large \bf Low-energy effective action \\[3mm] in  $5D, \cN=2$ supersymmetric gauge theory}
\\
\end{center}

\begin{center}

{\bf I.L. Buchbinder${}^{a,b,c}$, E.A. Ivanov$^c$ and I.B. Samsonov${}^{d}$
} \\
\vspace{5mm}

\footnotesize{ ${}^{a}${\it Department of Theoretical Physics,
Tomsk State Pedagogical University, 634061, Tomsk, Russia} \\
${}^{b}$ {\it National Research Tomsk State University, 634050, Tomsk, Russia}\\
${}^{c}$ {\it Bogoliubov Laboratory of Theoretical Physics, JINR, 141980 Dubna,
Moscow region, Russia}\\
${}^{d}$ {\it School of Physics, University of New South Wales, Sydney 2052, Australia
}}
\vspace{2mm}

{\tt joseph@tspu.edu.ru}, {\tt eivanov@theor.jinr.ru}, {\tt igor.samsonov@unsw.edu.au}
\end{center}

\vspace{2mm}

\begin{abstract}
We construct $\mathcal{N}=2$ supersymmetric low-energy effective
action of $5D, \mathcal{N}=2$ supersymmetric Yang--Mills (SYM)
theory in $5D, \mathcal{N}=1$ harmonic superspace. It is obtained
as a hypermultiplet completion of the leading $W \ln W$-term in
the $\mathcal{N}=1$ SYM low-energy effective action by invoking
the second implicit on-shell $\mathcal{N}=1$ supersymmetry. After
passing to components, the $\mathcal{N}=2$ effective action
constructed displays, along with other terms, the
$SO(5)$-invariant $F^{4}/X^{3}$ term. Though we specialize to the
case of $SU(2)$ gauge group spontaneously broken to $U(1)$, our
consideration is applicable to any gauge symmetry broken to some
abelian subgroup.
\end{abstract}

\vfill

\vfill
\end{titlepage}

\newpage
\renewcommand{\thefootnote}{\arabic{footnote}}
\setcounter{footnote}{0}




\section{Introduction}

Extended supersymmetry in diverse dimensions imposes stringent
constraints on the classical and effective quantum superfield
actions of gauge theories. The most prominent example is supplied
by the four-derivative term in the low-energy $4D, \mathcal{N}=4$
SYM effective action which, in the sector of $\mathcal{N}=2$ gauge
multiplet, is accommodated by a non-holomorphic superfield
potential~\cite{DS}. An $\mathcal{N}=4$ supersymmetric completion
of this potential was constructed in $\mathcal{N}=2$ harmonic
superspace \cite{BuIv,BIP,BBP,BP}, in $\mathcal{N}=3$ harmonic
superspace \cite{BISZ} and in various on-shell $\mathcal{N}=4$
harmonic superspaces \cite{BS,BS1} (see \cite{BIS-review} for a
review). By these works, the origin of non-renormalizability of
the $\mathcal{N}=4$ SYM low-energy effective action against
higher-loop quantum corrections was revealed and links with the
leading terms in the effective action of D3 brane on the
$AdS_{5}\times S^{5}$ background were established.

In $3D$ gauge theories, the constraints imposed by an extended
supersymmetry allowed one to determine the leading quantum
corrections in $\mathcal{N}=4$ SYM theory \cite{BPS09,BPS10}, and
to construct $\mathcal{N}=3$ superfield ABJM action \cite{N3ABJM}.
This technique proved also useful for displaying the structure of
the leading terms in $2D$ gauge theories with extended
supersymmetry \cite{S17}.

A common feature of all the above-mentioned results is that the
leading contributions to the low-energy effective actions in gauge
theories with extended supersymmetry take a reasonably simple form
upon the proper choice of the superfield description. The
constraints due to the extended supersymmetry (together with the
requirement of scale invariance) are strong enough to fix the form
of the superfield potentials in such theories up to an overall
coefficient to be found further from quantum considerations. In
some exceptional cases, e.g., in $4D, \mathcal{N}=4$ SYM theory,
the numerical value of this overall coefficient can yet be fixed
on the topological grounds, without the actual need to apply to
quantum computations \cite{BS}.

In this paper, we study the implications of extended supersymmetry
for the low-energy effective action of $5D$ SYM theory. This
theory is of interest from several points of view. It is
non-renormalizable by power-counting because of the dimensionful
coupling constant $g$, $[g]=-1/2$. Nevertheless, it was argued
that a non-perturbative quantum completion of this model describes
$6D, \mathcal{N}=(2,0)$ superconformal field theory compactified
on a circle \cite{Douglas,LPS,LPS1}. An additional confirmation of
this conjecture came from the exact computations of the partition
function in this theory by the localization technique
\cite{loc1,loc2,loc3,loc4,loc5}.

Despite the non-renormalizability of $5D, \mathcal{N}=1$ SYM, it
is still reasonable to study one-loop quantum corrections in this
theory, keeping in mind that in the odd-dimensional field theories
divergences can appear (within the dimensional regularization)
only at even loops. One-loop contributions to the effective action
of $5D, \mathcal{N}=1$ SYM theory were calculated in
Refs.~\cite{Kuz06,Pletnev} for the case of gauge group $SU(2)$
spontaneously broken to $U(1)$. The leading contribution is given
by the $5D$ supersymmetric Chern--Simons term \cite{Kuz06}, while
the next-to-leading one was found in \cite{Pletnev} in the form
%
\begin{equation}
\label{4deriv} c_{0}\int d^{5|8}z du \, W \ln \frac{W}{\Lambda
}\,,
\end{equation}
where $W$ is the $5D, \mathcal{N}=1$ abelian gauge superfield
strength, $\Lambda $ is a scale parameter, $[\Lambda ]=1$, and the
integration is performed over the full $\mathcal{N}=1$ harmonic
superspace with measure $d^{5|8}z du\equiv d^{5}x d^{8}\theta du$.
It is easy to check that the action (\ref{4deriv}) is $\Lambda
$-independent. The Chern--Simons term incorporates two-derivative
quantum corrections to the effective action, while (\ref{4deriv})
is $\mathcal{N}=1$ superfield extension of the four-derivative
``$F^{4}/\phi ^{3}$''-terms.

The purpose of this paper is to study leading terms in the
low-energy effective action of $5D, \mathcal{N}=2$ SYM theory in
harmonic superspace. Although such terms might be found by direct
quantum computations in $5D, \mathcal{N}=1$ superspace, we
determine them here on the symmetry grounds, just by constructing
$\mathcal{N}=2$ completion of the $5D, \mathcal{N}=1$ SYM
effective action via addition of the proper hypermultiplet terms.
The effective action constructed corresponds to the Coulomb branch
of $5D, \mathcal{N}=2$ SYM theory, with the gauge group being
broken to some abelian subgroup (e.g., the maximal torus), and, in
general, depends on the massless abelian $\mathcal{N}=2$ gauge
multiplets valued in the algebra of this subgroup. For simplicity,
we concentrate on the case of the gauge group $SU(2)$ and only
briefly discuss (in subsect. 2.3) the case of $SU(N)$ gauge
symmetry.

To start with, we point out that the $5D, \mathcal{N}=1$
Chern--Simons term does not admit an $\mathcal{N}=2$ completion
because it respects the invariance under $5D, \,\mathcal{N}=1$
superconformal algebra $F(4)$ which is unique and has no higher
$\mathcal{N}$ extensions \cite{Nahm}. According to this argument,
the Chern-Simons term is forbidden as a quantum correction in the
low-energy effective action of the $\mathcal{N}=2$ SYM theory.
From the field-theory point of view, it is possible to show, by
direct quantum computations in $5D, \mathcal{N}=1$ superspace,
that the two-derivative contributions (the Chern--Simons term) to
the $\mathcal{N}=2$ SYM effective action coming from the
hypermultiplet and from the ghost superfields cancel each other.
Indeed, the background field method for $5D,\,\mathcal{N}=1$ SYM
theory \cite{Pletnev} mimics the one for the $4D,\,\mathcal{N}=2$
SYM theory in harmonic superspace \cite{BFM,non-ren}. In
particular, the structure of ghost superfields in these theories
is the same. In Refs.~\cite{BK98,KM01} it was proved that the
hypermultiplet contributions to the $4D,\,\mathcal{N}=4$ SYM
one-loop effective action are fully canceled by the contributions
from the ghost superfields. Since this proof is purely formal, it
holds true for the $5D,\, \mathcal{N}=2$ SYM theory as well. Note
also that this cancellation is analogous to the well-known
phenomenon in $3D$ case \cite{BPS09,BPS10}, where Chern--Simons
term cannot arise as a quantum correction to the effective action
in supersymmetric gauge theories with $\mathcal{N}>2$.

As we demonstrate in the next section, the four-derivative term
(\ref{4deriv}), on the contrary, admits the unique hypermultiplet
completion under the requirement of $\mathcal{N}=2$ supersymmetry
involving an implicit $5D, \mathcal{N}=1$ on-shell supersymmetry
alongside with the manifest off-shell $\mathcal{N}=1$ one. The
procedure of constructing such a hypermultiplet completion is
quite analogous to the one developed in \cite{BuIv} for finding
the $\mathcal{N}=4$ hypermultiplet extension of the $4D,
\mathcal{N}=2$ SYM effective action. In the component formulation,
the $5D, \mathcal{N}=2$ effective action constructed displays the
$F^{4}/|X|^{3}$-term where $|X|^{2}$ is $SO(5)$-invariant bilinear
combination of scalar fields.

It is worth to point out that the term (\ref{4deriv})
(as well as its analogs for the higher-rank gauge groups) may
arise in quantum theory only as a one-loop quantum correction to
the effective action. Indeed, it is scale-invariant and so is
independent of the dimensionful gauge coupling constant $g$. On
the other hand, within the background field method in harmonic
superspace \cite{BFM,Pletnev}, all higher-loop Feynman graphs
involve a gauge superfield vertex with the coupling constant $g$.
Thus, all higher-loop quantum corrections to the effective action
cannot give rise to renormalization of the coefficient $c_0$ in
Eq.~(\ref{4deriv}) since they violate scale invariance.
However, in contrast to the $4D$ case, this coefficient is not
protected against non-perturbative corrections. Such corrections
will be discussed elsewhere.

Our last comment concerns the possible relation of the effective
action in $5D$ gauge theory to the D-brane low-energy dynamics.
The classical action of $5D, \mathcal{N}=2$ SYM theory with $U(N)$
gauge group can be interpreted as an action of a stack of $N$ D4
branes in flat space--time. The $\mathcal{N}=2$ supersymmetric
completion of the $F^{4}/\phi ^{3}$-term (\ref{4deriv})
can presumably be identified with that of the four-derivative term
in the low-energy effective action of a single D4 brane on the
$AdS_{6}\times S^{4}$ background.

\section{Low-energy effective action of $\mathcal{N}=2$ SYM theory}

In this section, we construct the low-energy effective action of
$\mathcal{N}=2$ SYM theory with the gauge group $SU(2)$ as a
hypermultiplet completion of the term (\ref{4deriv}). We
start our consideration with a brief account of the
$\mathcal{N}=1$ SYM and hypermultiplet models in $5D$ harmonic
superspace. We follow the notation and conventions of
Refs.~\cite{KL,Pletnev}.

\subsection{Classical action}

$\mathcal{N}=2$ gauge multiplet in $5D, \mathcal{N}=1$ harmonic
superspace is described by a pair of analytic superfields
$(V^{++},q ^{+}_{a})$, where $V^{++}$ is the $\mathcal{N}=1$ gauge
multiplet and $q^{+}_{a}\equiv (q^{+},-\bar{q}^{+})$ is the
hypermultiplet. The former is described by the classical action
written in the full harmonic superspace \cite{Zupnik87}
%
\begin{equation}
S_{\mathrm{YM }}= \frac{1}{2g^{2}} \sum _{n=2}^{\infty }
\frac{(-i)^{n}}{n} {\mathrm{tr}}\int d^{5|8}z du_{1} \ldots du_{n}
\frac{V ^{++}(z,u_{1})V^{++}(z,u_{2})\ldots
V^{++}(z,u_{n})}{(u^{+}_{1} u^{+} _{2}) (u^{+}_{2}
u^{+}_{3})\ldots (u^{+}_{n} u^{+}_{1})}\,, \label{Ssym}
\end{equation}
where $g$ is a coupling constant of mass-dimension $-1/2$. This
action yields the equation of motion
%
\begin{equation}
(\mathcal{D}^{+})^{2} W = 0\,, \label{DW}
\end{equation}
where $(\mathcal{D}^{+})^{2} \equiv {\mathcal{D}}^{+\hat{\alpha }}
{\mathcal{D}}^{+}_{\hat{\alpha }}$ and $W$ is a superfield
strength of the gauge $\mathcal{N}=1$ multiplet. It may be
expressed via the non-analytic prepotential $V^{--}$
%
\begin{equation}
W = \frac{i}{8} (\mathcal{D}^{+})^{2} V^{--}\,, \label{WV}
\end{equation}
which, in turn, is expressed through $V^{++}$ by the
harmonic-flatness condition
%
\begin{equation}
D^{++} V^{--} - D^{--} V^{++} + i[V^{++},V^{--}] =0\,.
\label{zero}
\end{equation}

The classical action of the hypermultiplet $q^{+a} (a=1,2)$ in the
adjoint representation of the gauge group reads
\cite{GIKOS,GIOS1,GIOS2}
%
\begin{equation}
S_{q} = \frac{1}{2g^{2}} {\mathrm{tr}}\int d\zeta ^{(-4)}
q^{+}_{a} {\mathcal{D}}^{++} q^{+a}\,, \label{Sq}
\end{equation}
where $d\zeta ^{(-4)}$ is the integration measure on the analytic
superspace and $\mathcal{D}^{++} =D^{++}+i[V^{++}$, is the
gauge-covariant harmonic derivative. The corresponding equation of
motion is
%
\begin{equation}
{\mathcal{D}}^{++} q^{+}_{a} =0\,. \label{Dq}
\end{equation}

The action of $\mathcal{N}=2$ gauge multiplet in $\mathcal{N}=1$
harmonic superspace is just the sum of (\ref{Ssym}) and (\ref{Sq}),
%
\begin{equation}
\label{S-class} S^{\mathcal{N}=2} = S_{\mathrm{YM }} + S_{q}\,.
\end{equation}
This action is invariant under the implicit $\mathcal{N}=1$
supersymmetry
%
\begin{equation}
\delta q^{+}_{a} = -\frac{1}{2} (D^{+})^{4} [\epsilon
_{a\hat{\alpha }} \theta ^{-\hat{\alpha }} V^{--}]\,, \qquad
\delta V^{++} = \epsilon ^{a}_{\hat{\alpha }} \theta
^{+\hat{\alpha }} q ^{+}_{a}\,, \label{hidden}
\end{equation}
where $\epsilon ^{a}_{\hat{\alpha }}$ is the relevant
anticommuting parameter. Though the equation (\ref{DW})
is modified for the total action (\ref{S-class}) by the
hypermultiplet source term in the right-hand-side, it is not the
case for the massless Cartan-subalgebra valued abelian superfields
which we will be interested in. In the abelian case, the equations
of motion for the $\mathcal{N}=1$ gauge multiplet
(\ref{DW}) and hypermultiplet (\ref{Dq}) are
simplified to the form
%
\begin{equation}
(D^{+})^{2} W = 0\,, \qquad D^{++} q^{+}_{a} = 0\,. \label{EOM}
\end{equation}
It is straightforward to show that on these equations the implicit
supersymmetry transformations (\ref{hidden}) are reduced
to
%
\begin{equation}
\delta q^{\pm }_{a} = \frac{i}{2} \epsilon _{a}^{\hat{\alpha }}
(D^{ \pm }_{\hat{\alpha }} W)\,, \qquad \delta W = -\frac{i}{8}
\epsilon ^{a}_{\hat{\alpha }} D^{- \hat{\alpha }} q^{+}_{a}
+\frac{i}{8} \epsilon ^{a}_{\hat{\alpha }} D ^{+\hat{\alpha }}
q^{-}_{a}\,. \label{hidden1}
\end{equation}

\subsection{$\mathcal{N}=2$ effective action}

The part of the superfield $\mathcal{N}=1$ SYM effective action
containing the component four-derivative term reads \cite{Pletnev}
%
\begin{equation}
\label{S0} S_{0} = c_{0} \int d^{5|8} z du \, W\ln
\frac{W}{\Lambda }\,,
\end{equation}
where $W$ is the abelian gauge superfield strength, $\Lambda $ is
a scale parameter and $c_{0}$ is a dimensionless constant. Owing
to the representation (\ref{WV}) implying $\int d^{5|8}
z du \, W = 0$, the action (\ref{S0}) is independent of
the scale $\Lambda $, $dS_{0}/d \Lambda =0$.

The precise value of the constant $c_{0}$ in the effective action
(\ref{S0}) depends on the gauge group representation
content of the hypermultiplet matter \cite{Pletnev}. Here, we do
not fix the value of this constant and construct $\mathcal{N}=2$
supersymmetric generalization of (\ref{S0}), keeping
$c_{0}$ arbitrary. This construction follows the same steps as in
Ref.~\cite{BuIv} where the similar $\mathcal{N}=4$ completion of
the leading term of the $4D, \mathcal{N}=2$ SYM effective action
was found.

The variation of the action (\ref{S0}) under the hidden
supersymmetry transformations (\ref{hidden1}) may be
cast in the form
%
\begin{equation}
\delta S_{0} = \frac{ic_{0}}{4} \int d^{5|8}z du \, \epsilon
^{a}_{ \hat{\alpha }} q^{+}_{a} \frac{D^{-\hat{\alpha }}W }{W}\,.
\label{S0-var}
\end{equation}
In deriving this equation we employed the abelian counterparts of
the relations (\ref{WV}), (\ref{zero}), the
equations of motion (\ref{EOM}) and integration by parts
with respect to the harmonic and covariant spinor derivatives.

The expression (\ref{S0-var}) may be partly canceled by
the variation of the action
%
\begin{equation}
S_{1} = c_{1} \int d^{5|8}z du \frac{q^{+a}q^{-}_{a}}{W}\,,
\end{equation}
where the coefficient $c_{1}$ will be defined below. The variation
of this action under (\ref{hidden1}) reads
%
\begin{equation}
\delta S_{1} = ic_{1} \int d^{5|8}z du \frac{q^{+a} \epsilon
^{\hat{\alpha }}_{a}(D^{-}_{\hat{\alpha }}W)}{W} -\frac{i}{8}
c_{1} \int d^{5|8}z du \frac{q^{+a}q^{-}_{a}}{W^{2}}(\epsilon
^{b}_{ \hat{\alpha }}D^{+\hat{\alpha }}q^{-}_{b} -\epsilon ^{b}_{
\hat{\alpha }}D^{-\hat{\alpha }}q^{+}_{b})\,. \label{deltaS1}
\end{equation}
The first term in the right-hand side in (\ref{deltaS1})
cancels the variation (\ref{S0}) if
%
\begin{equation}
c_{1} = -\frac{c_{0}}{4}\,, \label{c1}
\end{equation}
while the last term in (\ref{deltaS1}) may be cast in
the form
%
\begin{equation}
\delta (S_{0} + S_{1}) = -\frac{i c_{0}}{12} \int d^{5|8}z du
\frac{q ^{+a}q^{-}_{a}}{W^{3}} \epsilon ^{b}_{\hat{\alpha }}
q^{+}_{b} D^{- \hat{\alpha }}W\,.
\end{equation}
In deriving this expression, we integrated by parts and used
cyclic identities for $SU(2)$ indices. To cancel this expression
we need to add the next term
%
\begin{equation}
S_{2} = c_{2} \int d^{5|8}z du
\frac{(q^{+a}q^{-}_{a})^{2}}{W^{3}}\,, \qquad c_{2} =
\frac{c_{0}}{24}\,. \label{S2}
\end{equation}

Instead of evaluating the variation of the term
(\ref{S2}) we proceed to the general case and look for
the full $\mathcal{N}=2$ effective action in the form
%
\begin{equation}
S_{\mathrm{eff}}^{\mathcal{N}=2}= \int d^{5|8}z du\left [ c_{0} W
\ln \frac{W}{\Lambda }+\sum _{n=1}^{\infty }c_{n}
\frac{(q^{+a}q^{-} _{a})^{n}}{W^{2n-1}} \right ]\,, \label{Gamma}
\end{equation}
with some coefficients $c_{n}$. Let us consider two adjacent terms
in the sum in (\ref{Gamma}):
%
\begin{equation}
c_{n} \frac{(q^{+a}q^{-}_{a})^{n}}{W^{2n-1}} + c_{n+1}
\frac{(q^{+a}q ^{-}_{a})^{n+1}}{W^{2n+1}}\,.
\end{equation}
It is possible to show that the variation of the denominator in
the first term cancels the variation of the nominator in the
second term, if the coefficients are related as
%
\begin{equation}
(n+1) c_{n+1} = - c_{n} \frac{n(2n-1)}{n+2}\,.
\end{equation}
Taking into account Eq.~(\ref{c1}), we find from this
recurrence relation the generic coefficient
%
\begin{equation}
c_{n}= \frac{(-1)^{n}(2n-2)!}{n!(n+1)! 2^{n}}c_{0}\,. \label{cn}
\end{equation}
This allows us to sum up the series in (\ref{Gamma}) and
to represent the effective action in the closed form
%
\begin{equation}
S_{\mathrm{eff}}^{\mathcal{N}=2} = c_{0} \int d^{5|8}z du\, W\left
[ \ln \frac{W}{\Lambda }+ \frac{1}{2} H(Z) \right ], \label{Seff}
\end{equation}
where
%
\begin{equation}
Z\equiv \frac{q^{+a} q^{-}_{a}}{W^{2}} \,,
\end{equation}
and
%
\begin{equation}
H(Z) = 1+ 2\ln \frac{1 + \sqrt{1+2Z}}{2} +\frac{2}{3} \frac{1}{1 +
\sqrt{1+2Z}} -\frac{4}{3}\sqrt{1+2Z}\,. \label{H}
\end{equation}
It is easy to check that $H(0) =0\,, \;H^{\prime }(0) =
-\frac{1}{2}$, in agreement with (\ref{cn}).

The action (\ref{Seff}) is $\mathcal{N}=2$
supersymmetric extension of the effective action
(\ref{S0}). It would be interesting to reproduce this
result from the perturbative quantum computations in $5D$ harmonic
superspace, like it has been done in the $4D$ case in
\cite{BIP,BBP,BP}.

\subsection{Generalization to $SU(N)$ gauge group}

In the previous subsection we found the effective action
(\ref{Seff}) for a single massless $\mathcal{N}=2$ gauge
multiplet. Within a field theory, this effective action is
expected to come out from $ \mathcal{N}=2$ SYM theory
(\ref{S-class}) with the $SU(2)$ gauge group
spontaneously broken to its $U(1)$ subgroup. It is straightforward
to generalize this result to a higher-rank gauge group. For
instance, for the $SU(N)$ gauge group spontaneously broken to the
maximal torus $[U(1)]^{N-1}$ we obtain
%
\begin{equation}
S_{\mathrm{eff}}^{\mathcal{N}=2} = c_{0} \sum _{I<J}^{N} \int
d^{5|8}z du\, W_{IJ}\left [ \ln \frac{W_{IJ}}{\Lambda }+
\frac{1}{2} H(Z_{IJ}) \right ], \label{HSUn}
\end{equation}
where $Z_{IJ}= \frac{(q^{+a})_{IJ} (q^{-}_{a})_{IJ}}{W^{2}_{IJ}}$
and $W_{IJ}=W_{I} - W_{J}$, $(q^{\pm a})_{IJ}=q^{\pm a}_{I}-q^{\pm
a}_{J}$. The superfields $W_{I}$ and $q^{\pm a}_{I}$ obey the
constrains $\sum _{I} W_{I} = 0$, $\sum _{I} q^{\pm a}_{I} =0$ and
span the Cartan directions in the Lie algebra $su(N)$. The
function $H(Z_{IJ})$ for each argument $Z_{IJ}$ is given by the
same expression~(\ref{H}).

\subsection{Component structure}

We will be interested in deriving the term $F^{4}/X^{3}$ from the
effective action (\ref{Gamma}). To this end, it is
enough to leave only the following component fields in the
involved superfields:
%
\begin{eqnarray}
q^{+ 2} \equiv q^{+} &=&f^{i}(x)u^{+}_{i}\,, \qquad q^{+ 1} \equiv
\bar{q}^{+}=-\bar{f}^{i}(x) u^{+}_{i}\,,\nonumber\\
W&=& \sqrt{2}\phi
(x) - 2i \theta ^{+\hat{\alpha }}\theta ^{-\hat{\beta }}F
_{\hat{\alpha }\hat{\beta }}(x)\,. \label{Wcomp}
\end{eqnarray}
Here $\bar{\phi }= \phi $, $\overline{(f^{i})} = \bar{f}_{i}$ are
scalar fields and $F_{\hat{\alpha }\hat{\beta }} = F_{\hat{\beta }
\hat{\alpha }}$ is Maxwell field strength of the $\mathcal{N}=1$
gauge multiplet.

Substituting the superfield strength (\ref{Wcomp}) into
the first term in (\ref{Seff}), we find
%
\begin{equation}
S_{0} = c_{0}\frac{\sqrt{2}}{3}\int d^{5|8}z \frac{( \theta
^{+\hat{\alpha }}\theta ^{-\hat{\beta }}F_{\hat{\alpha }
\hat{\beta }})^{4}}{\phi ^{3}} =\frac{c_{0}}{4\sqrt{2}}\int
d^{5|8}z\frac{ \det F}{\phi ^{3}}(\theta ^{+})^{2}(\theta
^{+})^{2}(\theta ^{-})^{2}( \theta ^{-})^{2}\,, \label{S01}
\end{equation}
where $\det F = \frac{1}{4!} \varepsilon ^{\hat{\alpha }\hat{\beta
}\hat{\gamma }\hat{\delta }} \varepsilon ^{\hat{\mu }\hat{\nu
}\hat{\rho }\hat{\sigma }}F_{ \hat{\alpha }\hat{\mu }}
F_{\hat{\beta }\hat{\nu }}F_{\hat{\gamma } \hat{\rho
}}F_{\hat{\delta }\hat{\sigma }}$ and $(\theta ^{\pm })^{2} =
\theta ^{\pm \hat{\alpha }}\theta ^{\pm }_{\hat{\alpha }}$. We
integrate over the Grassmann variables according to the rule
%
\begin{equation}
\int d^{5|8}z \,(\theta ^{+})^{2}(\theta ^{+})^{2}(\theta
^{-})^{2}( \theta ^{-})^{2} f(x) = 4\int d^{5}x\,f(x)\,,
\end{equation}
for some $f(x)$. Thus the action (\ref{S01}) yields the
component term
%
\begin{equation}
S_{0} = \frac{c_{0}}{\sqrt{2}} \int d^{5}x \frac{\det F}{\phi
^{3}} \,. \label{S02}
\end{equation}

In a similar way we can perform the integration over the Grassmann
variables in the last term in (\ref{Seff}),
%
\begin{equation}
\int d^{5|8}z \,W H(Z) = \sqrt{2}\int d^{5}x \frac{\det F}{\phi
^{3}} [4 z^{4} H^{(4)}(z)+28 z^{3} H'''(z)+ 39 z ^{2} H''(z)+6z
H'(z)]\,, \label{2.29}
\end{equation}
where
%
\begin{equation}
z\equiv Z|_{\theta =0} = \frac{f^{i} \bar{f}_{i}}{2\phi ^{2}}\,.
\end{equation}
Substituting the function (\ref{H}) into
Eq.~(\ref{2.29}), we find
%
\begin{equation}
\frac{c_{0}}{2} \int d^{5|8}z \,W H(Z) = \frac{c_{0}}{\sqrt{2}}
\int d^{5}x \frac{\det F}{(\phi ^{2} + f^{i}\bar{f}_{i})^{3/2}}
-\frac{c _{0}}{\sqrt{2}} \int d^{5}x \frac{\det F}{\phi ^{3}}\,.
\end{equation}
The last term exactly cancels (\ref{S02}). As a result,
the total $F^{4}/X^{3}$ term in the component form of the
effective action (\ref{Seff}) is given by the expression
%
\begin{equation}
S_{\mathrm{eff}}^{\mathcal{N}=2} =\frac{c_{0}}{\sqrt{2}} \int
d^{5}x \frac{\det F}{|X|^{3}}+\ldots , \label{Scomp}
\end{equation}
where dots stand for the remaining terms and
%
\begin{equation}
\label{X} |X|=\sqrt{\phi ^{2} +f^{i} \bar{f}_{i}}\,.
\end{equation}
It is remarkable that the scalar fields appear in the denominator
in (\ref{Scomp}) just in the $SO(5)$ invariant
combination (\ref{X}). This is a non-trivial property,
since the field $\phi $ comes from the gauge $\mathcal{N}=1$
multiplet, while $f^{i}, \bar{f}_{i}$ from the hypermultiplet. In
the $SU(N)$ case (\ref{HSUn}), $S^{\mathcal{N}=2}
_{\mathrm{eff}}$ is given by a sum of the appropriate terms
(\ref{Scomp}).

\section{Summary and outlook}

In this paper, generalizing the approach of Ref.~\cite{BuIv} to
the $5D$ case, we constructed the leading term in the low-energy
effective action of $5D, \mathcal{N}=2$ SYM theory as the
appropriate sum of the effective action of $5D, \mathcal{N}=1$ SYM
theory and the interactions with the hypermultiplet. This
interaction is fixed, up to an overall coupling constant $c_{0}$,
by the requirement of the implicit on-shell $5D, \mathcal{N}=1$
supersymmetry extending the manifest off-shell $\mathcal{N}=1$
supersymmetry to an on-shell $5D, \mathcal{N}=2$ one. We discussed
in detail the case of the gauge group $SU(2)$ spontaneously broken
to $U(1)$, in which case the effective action depends on a pair of
single abelian $5D, \mathcal{N}=1$ gauge multiplet and
hypermultiplet, and then considered a more general situation with
the $SU(N)$ gauge group broken to its maximal torus, with $N-1$
pairs of such abelian multiplets.

The next obvious problem is to reproduce these effective actions
from the appropriate set of quantum $5D, \mathcal{N}=1$
supergraphs involving the interacting hypermultiplet and
$\mathcal{N}=1$ gauge superfields. Also, it would be interesting
to establish precise links with the relevant D-brane dynamics and
the $4D$ and $6D$ cousins of the $5D$ effective action
constructed. Finding out the explicit form of the next-to-leading
corrections to this effective action, based as well on the demand
of implicit $5D, \mathcal{N}=1$ supersymmetry, is another
interesting task for the future study. Finally, we expect that
results obtained can in principle be used to study the quantum
aspects of the twisted $5D,\ \mathcal{N}=2$ SYM theory known also
as the Witten-Haydys theory \cite{Witten11,Haydys} (see also
\cite{Zabzine}).

\subsection*{Acknowledgements} Authors are grateful to  RFBR grant No. 18-02-01046 for partial support.
The work of I.L.B and E.A.I was supported in part by the Ministry of Science and Higher Education
of Russian Federation, project No. 3.1386.2017.


\end{document}